\documentstyle[aps,prl,preprint,floats,epsfig]{revtex}

\begin{document}

\preprint{\tighten\vbox{\hbox{\hfil CLNS 98/1583}
                        \hbox{\hfil CLEO 98-14}
}}

\title{Search for Exclusive Rare Baryonic Decays of $B$ Mesons}

\author{CLEO Collaboration}
\date{\today}

\maketitle
\tighten

\begin{abstract}
We report the results of a search for the rare baryonic decay modes 
$B^{0} \to \Lambda \overline{\Lambda}$, $B^{+} \to \overline{\Lambda}p$, 
$B^{0} \to \overline{\Lambda} p \pi^{-}$, and $B^{0} \to p \overline{p}$ (and 
their charge conjugate states) using $5.8 \times 10^{6}$ $B \overline{B}$ 
pairs 
collected with the CLEO detector. We see no statistically 
significant signals in any of these modes and set 90\% confidence level  
upper limits on their branching fractions,
${\cal{B}}$($B^{0} \to \Lambda \overline{\Lambda}$) $< 3.9 \times 10^{-6}$, 
${\cal{B}}$($B^{+} \to \overline{\Lambda} p$) $< 2.6 \times 10^{-6}$, 
${\cal{B}}$($B^{0} \to \overline{\Lambda} p \pi^{-}$) $< 1.3 \times 10^{-5}$, 
and ${\cal{B}}$($B^{0} \to p \overline{p}$) $< 7.0 \times 10^{-6}$.
\end{abstract}
\newpage

{
\renewcommand{\thefootnote}{\fnsymbol{footnote}}

\begin{center}
T.~E.~Coan,$^{1}$ V.~Fadeyev,$^{1}$ I.~Korolkov,$^{1}$
Y.~Maravin,$^{1}$ I.~Narsky,$^{1}$ R.~Stroynowski,$^{1}$
J.~Ye,$^{1}$ T.~Wlodek,$^{1}$
M.~Artuso,$^{2}$ E.~Dambasuren,$^{2}$
% ...
\end{center}
 
\small
\begin{center}
$^{1}${Southern Methodist University, Dallas, Texas 75275}\\
$^{2}${Syracuse University, Syracuse, New York 13244}\\
% ...
\end{center}
\setcounter{footnote}{0}
}
\newpage


Evidence for the $b \to s$ quark transition, allowed in the Standard Model
by the penguin (internal W boson--quark loop) diagram, has been observed in 
both electromagnetic decays~\cite{btosg,kstarg} and hadronic decays of $B$ 
mesons to two mesons~\cite{kpi,etapr}. Charmless $B$ meson decays can also
arise through the $b \to u$ tree process as well as the 
Cabbibo suppressed $b \to d$
penguin process. These rare decay processes are of considerable
theoretical and experimental interest due to their importance in probing
the Cabibbo-Kobayashi-Maskawa (CKM)~\cite{CKM} picture of $CP$ violation 
within the Standard Model. 

\begin{figure}
\begin{center}
\epsfig{file=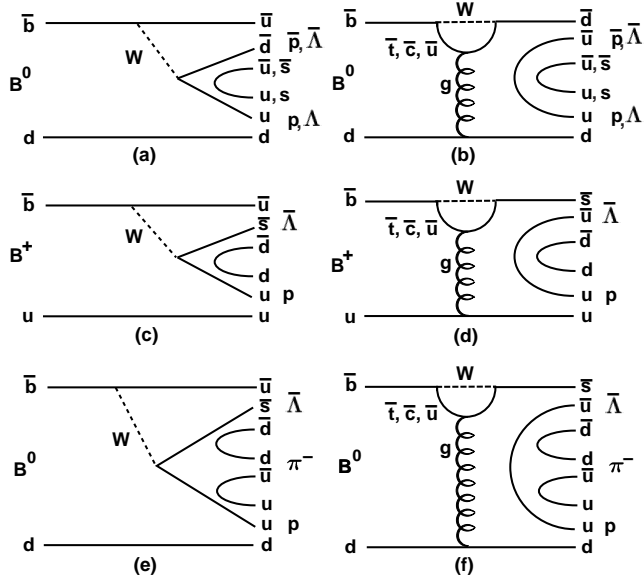, width=3.4in}
\end{center}
\caption{Tree (a,c,e) and penguin (b,d,f) decay processes which are expected 
to dominate the baryonic decays under consideration.}
\label{fig:feynman}
\end{figure}

This communication describes a search for charmless baryonic $B$ decays
to the final states $\Lambda \overline{\Lambda}$, $\overline{\Lambda}p$, 
$\overline{\Lambda} p \pi^{-}$, and $p \overline{p}$ (and 
their charge conjugate states). The dominant tree-level and one-loop penguin 
diagrams contributing to these decays are shown in Figure~\ref{fig:feynman}. 
The weak processes are similar to the meson decays
$B \to \pi \pi$, $K \pi$, and $KK$. We expect $B^{+} \to \overline{\Lambda} p$
and $B^{0} \to \overline{\Lambda} p \pi^{-}$ to be dominated by the $b \to s$
penguin diagram, while $B^{0} \to \Lambda \overline{\Lambda}$ and 
$B^{0} \to p \overline{p}$ should be dominated by the $b \to u$ tree
process. W-exchange, annihilation, penguin annihilation, and electroweak 
penguin processes can also contribute to these decay amplitudes~\cite{kohara}.
However, these diagrams are expected to have small contributions as 
compared to those shown in Figure~\ref{fig:feynman}. 
Theoretical predictions for 
these modes, scaled for a common value of $| V_{ub}| = 
0.0033$~\cite{PDG}, are given in Table~\ref{tab:results}. Both pole 
model~\cite{pole} and QCD sum rule calculations~\cite{chernyak} have been
performed. The pole model calculations yield higher predicted branching 
fractions for the rare baryonic modes. The search for the
three-body final state $\overline{\Lambda} p \pi^{-}$ is motivated by the 
experimental observation that ${\cal{B}}(B^{-} \to \Lambda_{c}^{+} 
\overline{p} \pi^{-})$/${\cal{B}}
(\overline{ B^{0}} \to \Lambda_{c}^{+} \overline{p})$ 
$> 1.0$ (90\% CL)~\cite{charmb}. We estimate the branching fraction for
$\overline{\Lambda} p \pi^{-}$ from the measured branching 
fraction of $B^{-} \to \Lambda_{c}^{+} \overline{p} \pi^{-}$, scaling
by the appropriate CKM and phase space factors. Penguin contributions could 
lead to significant enhancement over this rough estimate.


The data sets used in this analysis were collected with the 
CLEO~II~\cite{detector} detector at the Cornell 
Electron Storage Ring (CESR). It consists of $5.41~{\rm fb}^{-1}$ taken on the
$\Upsilon$(4S)
resonance and $2.79~{\rm fb}^{-1}$ taken below the $B\bar{B}$ production 
threshold.  The on-resonance sample contains 3.3~million $B\bar{B}$ pairs
taken before the installation of the silicon vertex detector,\cite{ross} 
(CLEO II data set) with the 
remainder of the 5.8~million $B\bar{B}$ pairs taken after the detector 
upgrade (CLEO II.V data set).

The momenta of charged particles are measured in a
67-layer tracking system operating
inside a 1.5 T superconducting solenoid.  The main drift chamber also
provides a measurement of the specific ionization loss, $dE/dx$,
which is used for particle identification.
Photons are detected by the 
7800-crystal CsI calorimeter. Muons are
identified using proportional counters placed at various depths in the
steel return yoke of the magnet.

Charged tracks are
required to pass track quality requirements based on the average hit residual
and the impact parameters in both the $r-\phi$ and $r-z$ planes.
Pairs of tracks with vertices displaced by at least $5$~mm from the primary
interaction point are taken as $\Lambda$ candidates. 
We require the $p \pi^-$ invariant mass to be within $10\ \rm{MeV/c^{2}}$ of 
the $\Lambda$ mass. Reconstructed $\Lambda$ baryons have a mass resolution of 
$1.4\ \rm{MeV/c^{2}}$. The impact parameter and $\Lambda$ flight distance 
requirements nearly eliminate feed-across between the 
$\Lambda \overline{\Lambda}$ and $\overline{\Lambda} p \pi^{-}$ final states.

Charged particles are identified as protons or pions from specific ionization
($dE/dx$) measurements from the drift chamber.
Electrons are rejected
based on their $dE/dx$ and the ratio of the track momentum to the associated 
shower energy in the CsI calorimeter.
We reject muons
by requiring that the tracks do not penetrate the steel absorber to a
depth greater than five nuclear interaction lengths. 
For the $p \overline{p}$
final state, a requirement on the time-of-flight of the protons and 
antiprotons is applied that eliminates 40\% of pion and 28\% of kaon 
background while retaining 89\% of the signal.


We calculate a beam-constrained $B$ mass 
$M_{BC} = \sqrt{E_{\rm b}^2 - {\bf p}_B^2}$, where ${\bf p}_B$ is the $B$
candidate momentum and $E_{\rm b}$ is the beam energy.
The $M_{BC}$ resolution is about 2.5~$\rm {MeV/c^2}$ for all modes.
We can capitalize on energy conservation if we define 
$\Delta E = \sum_{i} (E_i) - E_{\rm b}$, where $E_i$ are the
energies of the daughters that form the $B$ meson candidate.
The $\Delta E$ distribution is then centered about zero for reconstructed 
$B$ mesons and has a mode-dependent Gaussian width that ranges
from 17~MeV  
(for $\overline{\Lambda} p \pi^{-}$) to 25~MeV (for $p \overline{p}$).
We accept events with $M_{BC}$\ within $5.2-5.3$~$\rm {GeV/c^2}$\ and 
$|\Delta E|<200$~MeV. This region includes signal as well as a sideband used 
to fix the background normalization.

Backgrounds from $b\to c$\ as well as other $b\to u$\ and
$b\to s$\ decays are negligible for the two-body decay modes,
since the signal daughter particles are relatively light and are produced 
with high momentum. This is in contrast to the dominant $B$ meson decays which 
typically have a large number of final state particles with lower momenta. The 
decays $B^{0} \to K^{+} \pi^{-}$ and $B^{0} \to \pi^{+} \pi^{-}$ 
are displaced from zero by +287 and +330 MeV in $\Delta E$, respectively, and 
do not pass the sideband $\Delta E$ requirements for the $p \overline{p}$ 
final state. Possible 
backgrounds from 
$B$ decays to pseudoscalar-vector final states ($B\to PV$) 
such as $B\to \rho\pi$ and $B\to
K^*\pi$ are found in simulation studies 
to contribute at most $4\pm 4$ events to $p\overline{p}$. The simulations
are normalized to 90\% CL upper limits for ${\cal B}(B\to PV)$\cite{bigrare}. 
For the three-body mode $\overline{\Lambda} p \pi^{-}$, we
suppress backgrounds from $B$ meson decays as well as continuum by 
requiring that the $p$, $\pi^{-}$, and $\Lambda$ momentum be greater than 
0.70, 0.75, and 1.0~GeV/c, respectively. The placement of these cuts was 
determined by a signal squared over background optimization. These 
requirements reduce the background from other $B$ decays to less than 
5\% of the total background.

For this analysis, the main
background arises from $e^+e^-\to q\bar q$\ (where $q=u,d,s,c$).
Such events typically exhibit a two-jet structure and produce high 
momentum back-to-back tracks.
To reduce contamination from these events, we calculate the angle 
$\theta_{\rm{T}}$
between the thrust axis of the candidate tracks and the
thrust axis of the tracks and showers in the rest of the event. The 
distribution of $\cos\theta_{\rm{T}}$\ is strongly peaked at $\pm1$ for 
$q\bar q$\ events and is nearly flat for $B\bar B$\ 
events. For the low background final state $\Lambda \overline{\Lambda}$, we 
require
$|\cos\theta_{\rm{T}}|<0.95$ and for the remaining modes we require 
$|\cos\theta_{\rm{T}}|<0.9$.
A cut of $|\cos\theta_{\rm{T}}|<0.9$\ eliminates  $66\%$ of continuum 
background while retaining 87\% of the signal. 

A detailed GEANT-based Monte Carlo (MC) simulation~\cite{geant}
was used to determine the overall detection efficiencies 
(${\cal E}$) for each mode as shown in Table~\ref{tab:results}. 
Efficiencies contain branching fractions for $\Lambda \to p \pi^{-}$ where 
applicable. For the efficiency calculation we assume that the decay $B^{0} 
\to \overline{\Lambda} p \pi^{-}$ proceeds by phase space.~\cite{delta} 
From independent Monte Carlo samples we estimate systematic errors
on the efficiency determination; those errors are also given in
Table~\ref{tab:results}.

Additional discrimination between signal and $q\bar q$\ background is provided
by a Fisher discriminant technique, described in detail in 
Ref.~\cite{bigrare}.
The Fisher discriminant, ${\cal F}\equiv \sum_{i=1}^{11}\alpha_i y_i$, is a 
linear combination of 11 variables ($y_{i}$) where the coefficients 
($\alpha_i$) are chosen to maximize the separation between signal
and background Monte Carlo samples. 
The input variables are $|\cos\theta_{q}|$ (the cosine of the angle 
between the candidate thrust axis and the beam axis),  $|\cos\theta_B|$ 
(the cosine of the angle
between the $B$ meson momentum and the beam axis), and the energy and momentum
contained within nine concentric $10^\circ$ cones that surround the
candidate thrust axis. The sum of energy and momenta from tracks and
showers in the forward and backward cones are combined. The statistical 
separation between signal and continuum background afforded by the Fisher
discriminant is 1.5 standard deviations ($\sigma$) after events with
$|\cos\theta_{\rm{T}}|>0.9$ have been rejected.

\renewcommand{\arraystretch}{1.2}
\begin{table}
\begin{center}
\caption{Experimental results and theoretical 
predictions [8,9]. Branching fraction ($\cal{B}$) upper limits 
at 90\% CL are given in $10^{-6}$ units with (without) systematic errors
included. Quoted significance of the fit result is statistical only.}
\begin {tabular}{l c c c c c}
Mode&   $s$ & Sig. & ${\cal E}(\%)$ &${\cal{B}}$
& Theory ${\cal B}$\\
\hline
$\Lambda \overline{\Lambda}$ & $0.0^{+0.6}_{-0.0}$  & 0.0$\sigma$ & 
$14.8 \pm 1.8$ & $<3.9\ (3.4)$  & 0.13       \\
$p \overline{p}$   & $12.1^{+6.0}_{-4.9}$ & 2.8$\sigma$ & $48.8\pm2.7$
&$<7.0\ (6.0)$ & 0.6--4.8   \\

$\overline{\Lambda} p$ & $0.0^{+0.9}_{-0.0}$ & 0.0$\sigma$ & $29.3\pm2.8$
& $<2.6\ (2.3)$  & $\simeq$ 3        \\
$\overline{\Lambda} p \pi^{-}$ & $3.0^{+3.8}_{-2.9}$ & 1.1$\sigma$ 
&$14.6\pm1.7$ & $<13\ (11)$ & $\simeq$ 0.5 \\ 
\end {tabular}
\label{tab:results}
\end{center}
\end {table}
\renewcommand{\arraystretch}{1.0}
To determine the signal yield in each mode we perform an unbinned
maximum likelihood (ML) fit using $\Delta E$, $M_{BC}$, ${\cal F}$, $\Lambda$
and $\overline{\Lambda}$ masses (where applicable), 
and $dE/dx$ as input information for each candidate
event. Separate fits are performed for each mode. The likelihood function
is defined as
 \begin{equation}
{\cal{L}} = e^{-(s+b)}\prod_{i=1}^{N} \left[ sP_{S}(\vec{x_{i}}) + 
bP_{B}(\vec{x_{i}}) \right],
\label{like}
\end{equation}
where $s$ ($b$) are the signal (background) yields in the candidate sample (of
$N$ total events),
$P_{S}$ and $P_{B}$ are the signal and background probability density functions
(PDF), respectively, and $\vec{x_{i}}$ are the appropriate input variables
discussed above. The signal and background yields are varied until the 
likelihood function is maximized.
The probability density functions $P_{S}$ and $P_{B}$ are formed by the 
product of the best fit functional forms for signal and background 
distributions for each input variable. Signal MC is used to determine the
shape of the signal events and off-resonance data is used to determine the
background shapes. We use the sum of two Gaussians to parameterize the signal 
shapes for $\Delta E$, $M_{BC}$, and $\Lambda$ masses. Bifurcated Gaussians 
(with different low and high side widths) best 
describe the signal and background shapes for the Fisher discriminant and
a straight line is used to parameterize the $\Delta E$ background shape and
non-resonant $\Lambda$ mass background. When determining the background 
$M_{BC}$ distribution~\cite{argusbackground}
the kinematic endpoint of the off-resonance data
is matched to the on-resonance data by shifting the mass distribution by
$5.290\ {\rm{GeV}} - E_{\rm{b}}$. 

Systematic errors on the fitted yields are determined by varying the
PDFs used in the fit by their measurement errors. This variation is performed
with a Monte Carlo technique in which the PDF parameters are varied by  
Gaussian distributed random numbers. The width of the Gaussian is equal
to the measurement error of each parameter. For each variation the data
sample is  refit. The distribution
of the fit yield and upper limit yield from these MC experiments is fit
to a Gaussian whose width determines the ML fit systematic error. Important
correlations between PDF parameters are accounted for in this technique.

For each of the four rare baryonic decay modes, Table~\ref{tab:results} shows 
the signal yield ($s$) and statistical significance of the yield as determined
from the likelihood function. Since we see no statistically significant 
signals in any of the modes, we calculate $90\%$ confidence level 
upper limit yields ($s^{UL}$) by integrating the maximized likelihood function 
(${\cal L}_{\rm max}$) as follows:
\begin{equation}
{\int_0^{s^{UL}} {\cal L}_{\rm max} (N) dN
\over
\int_0^{\infty} {\cal L}_{\rm max} (N) dN}
= 0.90.
\label{eqn:upplim}
\end{equation}
The upper limit yield is then increased by
its systematic error (from the PDF variation) and the 
detection efficiency reduced by its systematic error before we calculate
the branching fraction upper limits shown in Table~\ref{tab:results}.
Table~\ref{tab:results} also gives the upper limits for each mode before
systematic errors have been included.

\begin{figure}[htbp]
\begin{center}
\epsfig{figure=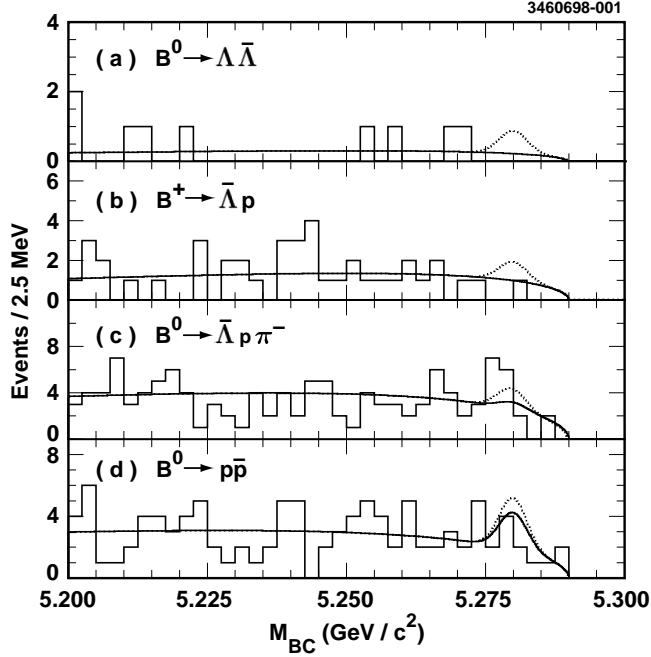,width=3.4in}
\end{center}
\caption{$M_{BC}$ projection plots for
(a) $B^0\rightarrow \Lambda \overline{\Lambda}$,
(b) $B^+\rightarrow \overline{\Lambda} p$, (c) $B^0\rightarrow 
\overline{\Lambda} p \pi^{-}$, and (d) $B^0\rightarrow p \overline{p}$.
The scaled projection of the total likelihood fit (solid curve)
and the fit reported upper limit (dotted curve) are overlaid.}
\label{fig:mass}

\end{figure}

\begin{figure}[htbp]
\begin{center}
\epsfig{figure=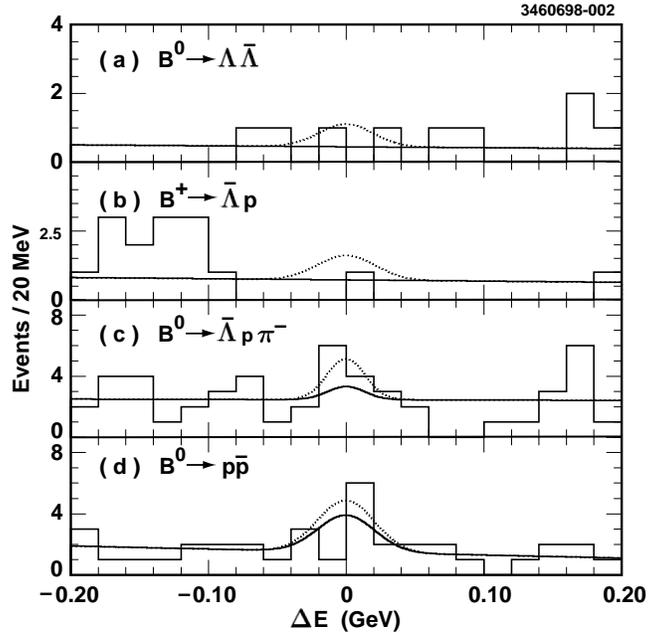,width=3.4in}
\end{center}
\caption{$\Delta E$ projection plots for
(a) $B^0\rightarrow \Lambda \overline{\Lambda}$,
(b) $B^+\rightarrow \overline{\Lambda} p$, (c) $B^0\rightarrow 
\overline{\Lambda} p \pi^{-}$, and (d) $B^0\rightarrow p \overline{p}$.
The scaled projection of the total likelihood fit (solid curve)
and the fit reported upper limit (dotted curve) are overlaid.}
\label{fig:de}
\end{figure}

Figures~\ref{fig:mass} and \ref{fig:de} show the $M_{BC}$ and $\Delta E$ 
projections, respectively, for each of the four modes. To reduce the 
background level in the plots we require $|\cos \theta_{\rm{T}}| < 0.8$ 
and apply a Fisher discriminant requirement, which eliminates roughly 80\% of 
the background and retains about 60\% of the signal. 
For the higher background modes ($\overline{\Lambda} p$, 
$\overline{\Lambda} p \pi^{-}$, and $p \overline{p}$) we require 
$|\Delta E| < 2.5\ \sigma_{\Delta E}$ for events to enter the $M_{BC}$ plot 
and $|M_{BC}-5.28| < 2.5\ \sigma_{M_{BC}}$ for events to enter the 
$\Delta E$ plot. Signal efficiency with these stricter requirements is about 
50\% of that quoted in Table~\ref{tab:results} while background efficiency is 
only 1-3 \%. Overlaid on these plots are the projections of the
PDFs used in the fit, normalized according to the fit results multiplied by 
the efficiency of the additional constraints. Both signal and signal upper 
limit yields are illustrated in the figures.


We have searched for decays of $B$ mesons to the baryonic final states
$\Lambda \overline{\Lambda}$, $\overline{\Lambda}p$, 
$\overline{\Lambda} p \pi^{-}$, and $p \overline{p}$ (and 
charge conjugate states) in 5.8~million $B \overline{B}$ pairs 
collected with the CLEO detector. 
We see no statistically significant evidence for signals in any of these
modes and set upper limits on their branching fractions. The upper limit for 
$B^{+} \to \overline{\Lambda} p$ is slightly lower than the theoretical 
estimate given in Reference~\cite{chernyak}. We see no evidence for
$b \to s$ penguin transitions (recently observed in decays to meson final
states) in $B$ decays to baryonic final states.

We gratefully acknowledge the effort of the CESR staff in providing us with
excellent luminosity and running conditions.
J.R. Patterson and I.P.J. Shipsey thank the NYI program of the NSF, 
M. Selen thanks the PFF program of the NSF, 
M. Selen and H. Yamamoto thank the OJI program of DOE, 
J.R. Patterson, K. Honscheid, M. Selen and V. Sharma 
thank the A.P. Sloan Foundation, 
M. Selen and V. Sharma thank Research Corporation, 
S. von Dombrowski thanks the Swiss National Science Foundation, 
and H. Schwarthoff thanks the Alexander von Humboldt Stiftung for support.  
This work was supported by the National Science Foundation, the
U.S. Department of Energy, and the Natural Sciences and Engineering Research 
Council of Canada.


\begin{references}
\bibitem{delta}
If the 
$\overline{\Lambda} p \pi^{-}$ final state were produced by the decay $B^{0}
\to \overline{\Lambda} \Delta^{0}$ with $\Delta^{0} \to p \pi^{-}$, then the
overall efficiency (including the $\Delta^{0} \to p \pi^{-}$ branching 
fraction) decreases to $1.9 \pm 0.3$\%.
\end{references}
\end{document}